\documentclass[journal]{IEEEtran}
\usepackage{all-packages}
\newcommand{\model}{\text{CANOA}}
\ifCLASSINFOpdf 
\else
\fi

\newcommand{\subparagraph}{}

\begin{document}

\title{CANOA: CAN Origin Authentication Through Power Side-Channel Monitoring}

\ifx\undefined\ANONYMIZED

\author{\IEEEauthorblockN{Shailja Thakur\IEEEauthorrefmark{1},
Carlos Moreno\IEEEauthorrefmark{2}, Sebastian Fischmeister\IEEEauthorrefmark{3}}\\
\IEEEauthorblockA{Electrical and Computer Engineering\\
University of Waterloo\\
Email: \IEEEauthorrefmark{1}s7thakur@uwaterloo.ca,
\IEEEauthorrefmark{2}cmoreno@uwaterloo.ca,
\IEEEauthorrefmark{3}sfischme@uwaterloo.ca}}


\else
\author{Anonymized Author(s)}
\fi

\maketitle

\acrodef{can}[CAN]{Controller Area Network}
\acrodef{ecu}[ECU]{Electronic Control Unit}
\acrodef{crc}[CRC]{Cyclic Redundancy Check}
\acrodef{dae}[DAE]{Denoising Autoencoder}
\acrodef{abs}[ABS]{Anti-Lock Brake System}
\acrodef{sdae}[SDAE]{Stacked Denoising Autoencoder}
\acrodef{dnn}[DNN]{Deep Neural Network}
\acrodef{pca}[PCA]{Principal Component Analysis}
\acrodef{svm}[SVM]{Support Vector Machine}
\acrodef{fft}[FFT]{Fast Fourier Transform}

\begin{abstract}
The lack of any sender authentication mechanism in place makes \ac{can} vulnerable to security threats. For instance, an attacker can impersonate an \ac{ecu} on the bus and send spoofed messages unobtrusively with the identifier of the impersonated \ac{ecu}. To address this problem, we propose a novel sender authentication technique that uses power consumption measurements of the \ac{ecu} to authenticate the sender of a message. When an ECU is transmitting, its power requirement is affected, and a characteristic pattern appears in its power consumption. Our technique exploits the power consumption of each ECU during the transmission of a message to determine whether the message actually originated from the purported sender. We evaluate our approach in both a lab setup and a real vehicle. We also evaluate our approach against factors that can impact the power consumption measurement of the \acp{ecu}. The results of the evaluation show that the proposed technique is applicable in a broad range of operating conditions with reasonable computational power requirements and attaining good accuracy.
  
\end{abstract}


%
\IEEEpeerreviewmaketitle

\newcommand{\probability}[1]{\mathrm{Pr}\left\{ #1 \right\}}
\newcommand{\fraction}[2]{\frac{\, #1 \,}{\, #2 \,}}
\newcommand{\Fraction}[2]{\displaystyle \fraction{#1}{#2}}
\newcommand{\norm}[1]{\left\lVert#1\right\rVert}
\def\inparenthesis#1{\left(#1\right)}
\newcommand{\vect}[1]{\boldsymbol{#1}}

\newtheorem{mydef}{Definition}

\def\implies{\Rightarrow}
\def\Implies{\Longrightarrow}
\def\iff{\Leftrightarrow}
\def\Iff{\Longleftrightarrow}
\def\assigned{~\leftarrow~}
\def\xor{\oplus}

\def\compromised{E_\mathrm{C}}
\def\target{E_\mathrm{T}}
\def\claimed{E_\mathrm{P}}
\def\predicted{\hat{E}}
\def\added{E_\mathrm{A}}

\newcommand{\bigOh}[1]{\mathrm{O}\left( #1 \right)}
\newcommand*\xbar[1]{%
   \hbox{%
     \vbox{%
       \hrule height 0.5pt 
       \kern0.5ex
       \hbox{%
         \kern-0.1em
         \ensuremath{#1}%
         \kern-0.1em
       }%
     }%
   }%
} 

\def\Section#1{Section~\ref{#1}}
\def\Figure#1{Figure~\ref{#1}}
\def\Eq#1{Equation~\eqref{#1}}
\def\Alg#1{Algorithm~\ref{#1}}
\def\Table#1{Table~\ref{#1}}
\def\Appendix#1{Appendix~\ref{#1}}
\ifx\undefined\FINAL

\def\proofingblankpage{\newpage}

\def\proofing#1{\textbf{\LARGE $\medstar$} \footnote{$\medstar\medstar$ \textbf{CM:}~~ #1 $\medstar\medstar$}}

\definecolor{Alarm}{rgb}{0.9,0,0}
\definecolor{Warning}{rgb}{0.8,0.3,0}
\definecolor{Good}{rgb}{0,0.75,0.25}

\def\brokensentence#1{{\color{Alarm} \proofing{Problem with sentence: #1}}}
\def\longsentence#1{{\color{Warning} \proofing{This sentence is too long; please rephrase}}}
\def\informal#1{{\color{Warning} \proofing{This sentence is too informal/coloquial: #1}}}
\def\praise#1{{\color{Good} \proofing{#1}}}
\def\unsubstantiated#1{{\color{Alarm} \proofing{Unsubstantiated claim: #1}}}
\def\incorrect#1{{\color{Alarm} \proofing{Factually incorrect: #1}}}
\def\careful#1{{\color{Warning} \proofing{Careful: #1}}}
\def\remove#1{{\color{Warning} \proofing{#1}}}
\def\change#1{{\color{Warning} \proofing{#1}}}
\def\warning#1{{\color{Warning} \proofing{#1}}}
\def\changed#1{{\color{Warning} \proofing{Changed it: #1}}}

\fi

%

\section{Introduction}
\label{sec:introduction}


\acresetall

\ac{can} is a communication protocol widely used in automotive systems for efficient real-time applications.
 However, its design exhibits significant security limitations. Among the most important of these limitations is the lack of sender authentication. Attackers can exploit any vulnerabilities in, e.g., the connectivity of automobile systems to infiltrate \acp{ecu} and inject spoofed messages on the network. Perhaps, the security of the \ac{can} protocol may not have been of such paramount importance back when the protocol came into existence in 1993. However, in recent decades security has become a critical aspect in automotive systems, given the increase in complexity and connectivity of modern vehicles~\cite{koscher-complexity-analysis}.

 Given the insecure nature of the \ac{can} bus, there has been a growing interest amongst the researchers to study the security of in-vehicle communication systems. For instance, Checkoway~et~al.~\cite{Checkoway:CEA} and Koscher~et~al.~\cite{koscher:ESA} demonstrate the potential vulnerabilities in automotive systems by studying and exploiting various attack vectors, including remote wireless connectivity such as Bluetooth, cellular, and DSRC . The work by Miller and Valasek~\cite{jeep-hack, more-jeep-hack} highlights the lack of authentication as one of the critical aspects of \ac{can} networks. They exploited the radio connectivity of the infotainment unit to hijack its functionality and send messages to other \acp{ecu}. The compromised unit impersonates \acp{ecu} involved in the control of critical physical attributes of the vehicle. This allowed them to demonstrate a remote attack that disrupts or hijacks the functionality of systems such as the engine, the brakes, and the steering. Moreover, dedicated websites have materialized providing procedures and guidelines for \ac{can} bus hacking and reverse engineering of vehicles~\cite{vehicle-reverse-engineering} for some of the major vehicle manufacturers.

\subsection{Problem Statement}

From the discussion above, it is evident that the inability of \ac{can} bus to authenticate the sender represents one of its most important security shortcomings.  This leads us to the problem that motivates our work, which can be stated as: given a message being transmitted on a \ac{can} bus, determine whether the message originated from the purported sender. Furthermore, if the message did not originate from the purported sender, then determine the actual sender of the message.

\subsection{Related Work}

Several approaches have been proposed in the past for sender authentication in the \ac{can} protocol. The solutions can be broadly categorized into two categories: message authentication using cryptographic techniques and authentication based on fingerprinting of physical characteristics of the transmissions.

Researchers have used traditional cryptographic techniques for message authentication~\cite{VHS11,LSV12,GM13} by including secret key as part of the \ac{can} frames to prevent forgery. However, the use of these techniques is restricted due to the limited size (8 bytes) of the \ac{can} frames and strict timing constraints in operation.

Fingerprinting techniques build upon side-channel analysis, or more in general, analysis of physical characteristics of the transmission. One such approach~\cite{Cho16} uses the clock skew of the periodically transmitted messages for intrusion detection. The method exploits the fact that the crystal clocks of the devices are not synchronized with each other resulting in a time deviation that is unique and stable over time and used as the \ac{ecu} identifier.
One of the limitations of the approach is that the technique does not work with aperiodic messages. Furthermore, the work by Sagong~et~al.~\cite{cloaking-clock} demonstrates that the method can be defeated by profiling and reproducing the timing patterns of the target \ac{ecu}.

Murvay and Groza~\cite{MG14} devised a fingerprinting approach that uses voltage variations to fingerprint the ECUs for sender authentication. This approach applies only to the voltage measured on a low-speed \ac{can} bus, while vehicles today operate at varying speeds from 10\,Kbps to 1\,Mbps depending on complexity and functionality. To overcome this limitation, Choi~et~al.~\cite{Choi16} proposed an approach that generates \ac{ecu} fingerprints from voltage measurement using both time and frequency domain features and used a supervised classification algorithm for sender authentication. Although the approach detected transmitter with improved accuracy, the method has a practical limitation that the measurements are collected at an extremely high sampling rate (2.5\,Gsps), and it works with a fixed message format.

Cho~et~al.~\cite{viden} developed a model for sender identification by fingerprinting the \acp{ecu} using voltage measurements against dominant bits of the transmissions.
Kneib and Huth proposed Scission~\cite{scission}, an \ac{ecu} profiling technique that builds upon the idea of Viden for sender authentication.
Scission relies on the use of all the transmission bits instead of just the dominant bits to construct \ac{ecu} profile.
Similarly to~\cite{Cho16}, this technique relies on physical characteristics that conceivably could be profiled and imitated by a different device. Moreover, they could potentially be affected through access to subsystems outside the device implementing the technique, as shown by~\cite{attack-viden}.  We do acknowledge that any such attacks would require temporary physical access to the target \ac{can}, to add a custom device on the network.

Two recent works~\cite{biangulation,tcan} propose techniques for sender authentication on a \ac{can} network based on similar underlying ideas: determine the physical location of the sender based on the propagation time of the signal as it travels through the \ac{can} bus. Though the techniques do exhibit some important advantages, the works are feasibility studies. Moreover, the techniques still suffer the limitation that an attacker can potentially defeat it with temporary physical access to the target \ac{can} bus.

\subsection{Contribution}
In this work, we propose \model{}, a novel technique for sender authentication using power consumption measurements of \acp{ecu} as identifying characteristics. Power consumption measurements leak relevant and critical information about the sequence of operations executed in the \acp{ecu}~\cite{Kocher:DPA}.
Specifically, we exploit the correlation between the power consumption of each \ac{ecu} and its state (transmitting or not transmitting).  From the power consumption measurements for all the \acp{ecu}, \model{} determines the actual sender when a transmission is observed on the bus (with a purported sender in its data), which constitutes an effective sender authentication mechanism. One key and unique advantage of \model{} is that the classification is based on physical characteristics of the transmitting \ac{ecu} that are guaranteed to be non-clonable. We can see that this is the case, given the strict relationship between \acp{ecu} activity (in particular, transmitting vs. not transmitting) and their power consumption patterns: if an \ac{ecu} $E$ is not transmitting, it is \emph{physically impossible} for another \ac{ecu} to make the power consumption of $E$ exhibit the same pattern it does when it is transmitting. We observe that one condition for our technique to be effective is that the power consumption patterns when transmitting and not transmitting, and even while receiving, must not only be different: the difference should be large enough for the patterns to be distinguishable. This is one aspect that the results of this study confirm.

\vskip \medskipamount
The contributions of this paper are as follows:
\begin{itemize}
	\item We propose and implement \model{}, a technique for sender authentication using power consumption characteristics of transmitting and non-transmitting states of the \acp{ecu}.


    \item We show the applicability of \model{} in practical settings by evaluating our proposed approach for sender authentication in a lab setup and a real vehicle.

    \item Lastly, we demonstrate the feasibility and technical viability of \model{} by studying the impact of the variations in bus speed, message format and source code on the accuracy of the \model{}.
\end{itemize}

\subsection{Organization of the Paper}
The remainder of the paper is organized as follows: \Section{sec:background} gives a background on the related concepts followed by proposed technique in \Section{sec:proposed-technique}. \Section{sec:exp-evaluation} describes the setup and results of experimental evaluation, followed by a brief discussion and suggested future work in \Section{sec:discussion}.

\section{Background}
\label{sec:background}

This section provides some background on concepts related to our work.  We briefly describe some aspects of the \ac{can} bus that is used in automotive systems, power-based program tracing or monitoring, and machine learning-based models.

\acresetall

\subsection{Controller Area Network}
\label{subsec:can}

\ac{can} uses a broadcast topology where multiple nodes (\acp{ecu}) can connect and exchange data~\cite{CAN}. The physical layer of \ac{can} is a twisted-pair cable for serial communication using differential signalling. The operation of the \ac{can} bus at the physical layer is based on ``open-collector'' or ``open-drain'' connected devices, implementing a ``wired AND'' connection. Any device can, without causing any conflict or short-circuit on the bus, assert a logical 0 on the bus, independently of what any other module is transmitting. Releasing the bus implicitly brings it to a logical 1, provided that no other device is asserting a logical 0.

\begin{figure}[h]
\centering
    \includegraphics[width=0.9\linewidth]{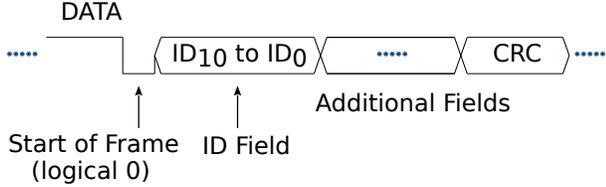}
    \caption{Simplified diagram of a \ac{can} frame.}
    \label{fig:can-frame}
\end{figure}

The devices themselves arbitrate access to the bus. To this end, a priority field is used, the \emph{ID} field, as illustrated in \Figure{fig:can-frame}. The figure and our discussion are limited to the 11-bit base frame, which is the most commonly used \ac{can} frame format. However, we emphasize that our proposed technique operates equally effectively with either 11-bit IDs or with the extended frame 29-bit IDs. Lower values for this ID represent higher priority, and devices read back the state of the bus to detect collisions: if a device transmitting a 1 as part of the ID field reads the bus and observes a logical 0, then it concludes that some other module of higher priority is transmitting, so it releases the bus. This is the case since the ID is transmitted MSB to LSB.

Additional issues related to the devices' reaction to collisions are not relevant to our work, so we omit any further details.
~Every \ac{can} frame includes a 15-bit \ac{crc} field for fault-tolerance \hbox{purposes\,---\,that} is, to protect the
transmission from unintentional errors due to noise or other artifacts at the hardware level. These may include
loose wires, defective or aged electronics, etc.

\subsection{Power-Based Program Tracing or Monitoring}

Based upon the field of side-channel analysis~\cite{Kocher:DPA}, power-based program tracing or monitoring have appeared in the literature in recent years~\cite{side-channel-disassembler,moreno:lctes13,WattsupDoc,b-side,ccs16}. Among these, the earlier works~(\cite{side-channel-disassembler,moreno:lctes13}) propose the technique as a means to reconstruct the program's execution trace in a deployed (uninstrumented) embedded device, although both works mention other uses. \cite{WattsupDoc,b-side,ccs16} focus on observing the power consumption of an embedded device to detect security attacks, following the rationale that such attacks would cause the device to deviate from its normal operation.  Moreno and Fischmeister argue that this is an effective technique for monitoring safety-critical embedded systems to enforce both safety and security properties~\cite{who-watches-the-watchers}.

The basic idea is to exploit the relationship between a device's power consumption and its operation (more precisely, what a processor is executing). By monitoring power consumption, one can detect deviations from normal operation. This can be done either by attempting an explicit reconstruction of the program's execution trace or simply through profiling the power consumption patterns during normal operation and detecting deviations from the profiled normal behaviour.

In our proposed technique, the use of power consumption monitoring has a slightly different, yet closely related goal: we are interested in exploiting the correlation between the power consumption of \acp{ecu} and one aspect of the \acp{ecu} program's \hbox{execution\,---\,whether} the given \ac{ecu} is transmitting on the \ac{can} bus. Though this leads to a more limited accomplishment in terms of enforcing security, it has two important advantages: (1)~Our technique can detect attacks executed by a device added to the system by an attacker with physical access (we will discuss this aspect in more detail in \Section{subsec:attack-model}); and (2)~we can achieve a significantly higher accuracy compared to existing power-based monitoring techniques (or similar accuracy at much lower computational/processing power requirements). This is the case because our system only needs to reconstruct a feature that represents a much lower amount of
information, compared to reconstructing the complete execution trace or detecting minor deviations (or deviations during a short amount of time) in the power consumption.

\subsection{Machine Learning Based Classification}
\label{subsec:deep-NN}

Classical supervised machine learning approach relies on statistical pattern recognition to perform classification tasks. Given a set of input observations $\{x_1, x_2, ~\cdots~ , x_R\} \subseteq \mathcal{X}$ and a set of output class labels $\mathcal{C}=\{C_1, C_2, ~\cdots~ , C_{S}\}$, the goal of any statistical learning method is to learn a mapping $\hat{f}$ from input observations to output class labels, $\hat{f}: \mathcal{X} \rightarrow \mathcal{C}$. This mapping is an estimate of the true but unknown function $f$ that maps each input observation $x$ to the class to which it corresponds.

Typical classifier implementations rely on features extracted from the observation $x \in \mathbb{R}^n$ for the operation where $n$ is the input dimension.
For many applications, however, it is difficult to identify the relevant features because of the complicated nature of observations. One solution to this problem is to use a dimensionality reduction technique to extract relevant features in lower-dimensional space. \ac{pca}~\cite{statisticalearning} is a dimensionality reduction technique that seeks to retain maximum variability in the input by projecting the input to linear subspace. Features from the linear subspace can be fed as input to a classifier and perform the classification task. In our proposed technique, we use a \ac{svm}~\cite{statisticalearning} classifier for binary classification tasks. \ac{svm} is a discriminative classification algorithm best suited for high dimensional inputs. In this paper, we aim to detect the state of an \ac{ecu} during transmission by classifying a segment of power consumption measurement of the \ac{ecu} into the class of transmission and non-transmission. And, a simple machine learning algorithm such as an \ac{svm} is sufficient to serve the purpose.

\section{Proposed Technique}
\label{sec:proposed-technique}

This section presents the proposed technique \model{} for sender authentication on a \ac{can} network. We first describe the attack model and assumptions, followed by the proposed approach.

\subsection{Attack Model and Assumptions}
\label{subsec:attack-model}



In our attack model, a target \ac{ecu} is denoted $\target$, and it represents the \ac{ecu} that the attacker wants to impersonate. That is, the attacker's goal is to transmit messages with IDs that correspond to $\target$.  The purpose of such an attack may be to cause some other \ac{ecu} to operate on false \hbox{data\,---\,data} that is logically correct but with contents that are under the attacker's control. For example, $\target$ could be an \ac{ecu} connected to sensors, and an attacker could transmit false temperature or speed data that could result in physical damage to the transmission or the engine. Notice that this implies that $\target$ does not have any vulnerabilities that the attacker can exploit; for example, an \ac{ecu} without any connectivity to the Internet or mobile networks, when we consider remote attackers.

\begin{figure}[h]
    \centering
    \includegraphics[width=0.7\linewidth]{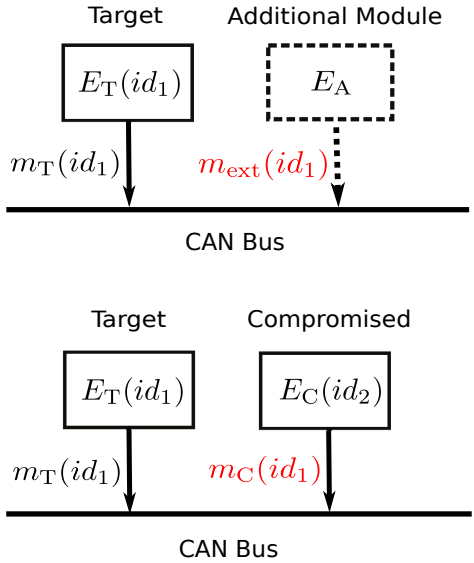}
    \caption{Examples of attack scenarios on a \ac{can} network. The dashed box in the top figure shows the additional illegitimate module attached to the network and boxes with solid lines are the legitimate \acp{ecu} on the network.}
    \label{fig:attack-scenario}
\end{figure}

An attacker is capable of sending crafted messages using the ID of the $\target$ in the following scenarios:
\begin{itemize}
\item \textbf{Compromised \ac{ecu}: }
In this scenario, an attacker gains access to an \ac{ecu}, denoted $\compromised$. Such compromised \ac{ecu} may be, for example, one with connectivity that exposes some vulnerabilities to remote attackers, or exposes services (e.g., open ports) that the designers of the vehicle did not intend to offer (and are unaware that those services are active and available).

\item \textbf{Added \ac{ecu}:}
Our attack model includes an adversary that may temporarily gain physical access to the vehicle and the target \ac{can} bus. Thus, they can attach an additional module, denoted $\added$, capable of transmitting \ac{can} messages to the existing network.  This additional module can be any arbitrary, custom hardware with firmware entirely created by the attacker, thus capable of listening and transmitting without any restrictions.
\end{itemize}
\Figure{fig:attack-scenario} illustrates the above two attack vectors.


Our attack model also includes the possibility of an attacker tampering with ongoing transmissions.
In particular, the malicious \ac{ecu} (be it $\compromised$ or $\added$) can hijack an ongoing transmission from \ac{ecu} $E_1$ to change its ID and make it look like the sender is \ac{ecu} $E_2$. ~We observe that this is feasible given the ``wired AND'' nature of the \ac{can} bus, allowing an attacker to change any 1's to 0's in an ongoing transmission. ~$E_1$ will determine that a higher-priority frame is being transmitted, and will withdraw from the bus; from that point, the attacker can complete the transmission.  ~Although this attack does not affect \model{}'s ability to determine that the sender is not $E_2$, it may be beneficial for the attacker to attempt to avoid detection by shifting the blame to $E_2$.

\vskip \medskipamount
\noindent\textbf{Assumptions}
\vskip \smallskipamount

Our proposed technique operates under the following assumptions and limitations:
\begin{itemize}
\item \textbf{Secure \model{} Implementation:} ~We assume that our proposed technique is implemented in a secure and tamper-proof manner.
In principle, we would expect the module implementing our technique to be physically isolated from any \acp{ecu} on the target \ac{can} bus.

\item \textbf{No DoS:}
~We assume that an attacker will not perform a ``brute'' denial-of-service attack
on the \ac{can} bus. In particular, if the attackers can place an arbitrary device, they
can certainly disrupt every transmission or even assert a permanent logical 0
on the bus, effectively severing all communications between any \acp{ecu} on
that bus.  Though this sort of ``trivial'' attack seems powerful, it will
also be trivial to detect; the design of many vehicles perhaps already
includes safety mechanisms that would appropriately deal with situations like
this~\cite{iso26262}.

\item \textbf{Types of Attacks:} Our proposed technique \model{} concentrates on impersonation attacks. If a transmission from an \ac{ecu}
$\compromised$ contains malicious data but legitimate ID (i.e., an ID that does correspond to $\compromised$), then our system will not
flag any anomalies or suspected attacks. Furthermore, our method does not detect if a new source address appears on the network; however, this is an easy intrusion to detect.

\end{itemize}

 \subsection{Proposed Approach}
In this section, we explain the technique to verify the sender of a message given the power consumption measurements of the \acp{ecu} on the bus.
Let $\mathcal{E} = \{E_1, E_2, ~\cdots~, E_K\}$ be a set of $K$ \acp{ecu}. Let $\mathcal{S} = \{S_1, S_2, ~\cdots~, S_K\}$ be a set of $K$ source addresses. A source address is unique to a sender; however, an \ac{ecu} may generate messages with more than one source address where the source address of a message can be derived from the ID of the message.~Let $\bm{P}_k \in \mathbb{R}^L$ represent a vector of power consumption measurements of length $L$ from \ac{ecu} $E_k$. Let a decoded transmission be represented by a tuple $(t, S)$ where $t$ represents the start time of the transmission, and $S$ denotes the source address of the transmission. Let a set of $K$ classification models be denoted by $\mathcal{F} = \{f_1, f_2, \cdots, f_K\}$.
Let an estimate of the transmission window be denoted by $\tau$, corresponding to the amount of time during which transmission is observed on the bus. We will use the term \emph{power trace} to refer to a segment of power consumption measurement of length $\tau$. Notice that a power trace may contain power consumption measurements during a time interval where the ECU was not transmitting.



\subsubsection{Generating features of transmissions}
\label{subsec:features-transmissions}
   In this section, we describe the method to construct features of transmissions using power consumption measurements of the \acp{ecu}.
   As the first step towards the implementation of \model{}, we capture analog voltage signal and power consumption measurements of the \acp{ecu}. From the captured voltage signal, we decode the times of occurrences and source addresses of the transmissions. We also calculate an estimate of the transmission window ($\tau$), which is equal to the mean of transmission windows corresponding to $N$ decoded transmissions. Using the power consumption measurements from the \acp{ecu} and the decoded transmissions, we construct features of the decoded transmissions, which is used as training data for the \ac{ecu} state identification model implementation.

   \Alg{alg:fingerprints} summarizes the steps for the construction of the feature vectors from transmissions using power consumption measurements of the \acp{ecu}. The input to the algorithm is a set of power consumption measurements from the \acp{ecu}, $\mathcal{P}$, a sequence of decoded transmissions $\mathcal{T}$. The output generated by the algorithm is a set of pairs $\{(\bm{X}_1, \bm{y}_1), (\bm{X}_2, \bm{y}_2), \cdots, (\bm{X}_K, \bm{y}_K)\}$ where $\bm{X}_k \in \mathbb{R}^{N \times M}$ is a matrix of $N$ feature vectors of length $M$ and $\bm{y}_k$ is an array of the source addresses corresponding to the transmissions for \ac{ecu} $E_k$. As a preprocessing step, we first normalize the power consumption measurements from the \acp{ecu}. This is done to unify the scale of the power consumption measurements from different \acp{ecu} to a common scale. In the algorithm, in the expression for normalization, $\xbar{\bm{P}}_k$ and $s_k$ are the estimates of mean and the standard deviation of a sample of power signal from the \ac{ecu} $E_k$. The normalization of power signals is followed by generating features from transmissions as follows: For a transmission $(t, S)$ and an \ac{ecu}, $E_k$, fetch the power trace $\bm{P}_k[\tau]$ corresponding to the \acp{ecu} $E_k$ starting with the time $t$ for the length of the transmission window $\tau$. To the extracted segment of power signal, we apply a windowing called Tukey window~\cite{fourier-analysis}. The Tukey window method helps reduce the amplitude of discontinuities at the boundaries of the segment by multiplying it with a finite-legth window with an amplitude that varies smoothly and gradually toward zero at the edges. This is followed by applying \ac{fft} to the windowed segment of power-trace. This helps recognize and eliminate the frequency componenets that are predominantly noise. This is followed by computing the \ac{pca} of the resultant power trace in the frequency domain to filter out the frequency compoennts which posses most of the variance in the segment, and concatenating the first $M$ principal components of the \ac{pca}, which form the feature vector $\bm{x}$ for the transmission $(t, S)$ with the decoded source address $y$ as the label.


    \begin{algorithm}
       \begin{algorithmic}[1]
         \Require{$\mathcal{P} \in \mathbb{R}^{K \times L}$, $\mathcal{T}$: a set $N$ of transmissions, $\mathcal{E}$: a set of $K$ \acp{ecu}, $\tau$: transmission window}
         \Function{powertraces}{$ \mathcal{T}, \mathcal{P}, \mathcal{E}$}


        \State $\mathcal{X} \gets 0$ \Comment{Initialize a set of $K$ $N \times M$ matrices}

        \State $\mathcal{Y} \gets 0$ \Comment{Initialize a set of $K$ $N$-dimensional vectors }

         \State $\mathcal{P}^* \gets \Fraction{\bm{P}_k - \xbar{\bm{P}}_k}{s_k}, 1 \leqslant k \leqslant K$ \Comment{Normalize $\mathcal{P}$}

         \ForAll{$E_k \in \mathcal{E}$}

         \For{$n \gets 1$ to $N$}      \Comment{Iterate $\mathcal{T}$}

           \State $\bm{x} \gets \bm{P}^*_k[t  : (t  +  \tau)]$
           \State $\bm{x} \gets \text{Apply}\_\text{Tukey}\_\text{Window}(\bm{x})$
           \State $\bm{x} \gets \text{Apply}\_\text{FFT}(\bm{x})$
           \State $\bm{x} \gets \text{Apply}\_\text{PCA}(\bm{x})$
           \State $\bm{X}_k[n] \gets \bm{x}[0:M]$
           \State $\bm{y}_k[n] \gets S$


         \EndFor
         \EndFor
         \State \Return{($\mathcal{X}, \mathcal{Y}$)}
         \EndFunction
         \end{algorithmic}
         \caption{{Generate Power Traces}}
         \label{alg:fingerprints}
       \end{algorithm}

\subsubsection{Model Implementation}
\label{subsec:classification}

   To perform sender authentication, we implement a set of binary (transmission/non-transmission) classification models separately for all the source addresses observed on the \ac{can} bus. As a source address uniquely identifies an \ac{ecu} on the bus, the model per source address acts as a sender state identifier, and hence, a sender authenticator. Even if multiple source addresses are associated with an \ac{ecu}, the fact that only one \ac{ecu} transmits at a time, thus, the non-overlapping features of transmission and non-transmission for the \acp{ecu} helps in identifying the correct state of the \ac{ecu}. The model implementation is subdivided into two stages: training and classification.


   \begin{itemize}
  \item \textbf{Training:} For every ($E_k$, $S_k$) where $S_k$ is mapped to the \ac{ecu} $E_k$, we train a classifier $f_k:\bm{X}_k \to \mathbb{R}^2$ that maps the set of features of transmission to a vector of probabilities, which signifies the strength of the classifier in the class of transmission (1) and non-transmission (0). The training data $(\bm{X}_k, \bm{y}_k)$ comprises of a labelled set of extracted features of trasnmissions from $E_k$ where an example $\bm{x} \in \bm{X}_k$ is labelled as one (or, to the class of transmission) if the trasnmission was observed with source address $S_k$; otherwise, the example is labelled as zero (or, to the class of non-transmission because the transmission is observed with the source address from the set  ${\mathcal{S} \setminus S_k}$ which are not mapped to $E_k$). In other words, the class of transmission contains features of transmissions with the source address $S_k$ from the \ac{ecu} $E_k$; whereas, the class of non-transmission contains features of transmissions with other source addresses $\mathcal{S} \setminus S_k$ from the \ac{ecu} $E_k$. Using the prepared training set, we train the model until convergence; that is until the error between the predicted and the true class approaches a specific threshold, $\epsilon$. At convergence, \model{} generates a trained classifier, $f_k$, which can be used for predicting the probability of transmission from source address $S_k$.

  \par

   \item \textbf{Classification:} As transmission is observed with only one source address on \ac{can} bus, only one model $f_k \in \bm{F}$ from amongst the models $\bm{F}$ will report the highest probability of transmission. The $(E_k, S_k)$ corresponding to the model $f_k$ that reports the highest probability of transmission is the actual source of the transmission. During classification, for every new decoded transmission $(t, S)$, \model{} determines the transmitting state of each of the \ac{ecu} $E_k \in \mathcal{E}$ by feeding the features of transmission calculated at time $t$ to the models $f_k \in \bm{F}$.
    We apply an activation function called softmax to the vector of the probability of transmissions from the models so that the output vector sums to one, a property essential for retaining the probabilistic nature of the output. Based on the prediction probabilities of the models, an outcome of 1 is assigned to the source address $S_k$ for which the corresponding model's likelihood of transmission is greater than a predefined threshold $\delta$, and is greater than that of all the other model's predictions. And an outcome of 0 is assigned for all the other source addresses $\mathcal{S} \setminus S_k$.

   \end{itemize}

   \subsection{Attack Detection}
   \label{subsec:attack-detection}

 In this section, we describe how to use the model classification result to detect the attack give the attack model explained in \Section{subsec:attack-model}. Given a transmission $T = (t, S)$ with the start time of transmission $t$ and $S$ as the source address of the message in transmission, \model{} identifies the purported sender $E_\mathrm{P}$ of the transmission, and uses the power trace of $E_\mathrm{P}$ as input to the classifier to determine whether $E_\mathrm{P}$ is the source of transmission. Based on the prediction, \model{} determines whether the transmission constitutes an attempted impersonation attack.

\begin{itemize}
    \item \textbf{Detection of Impersonation Attacks:} Consider a transmission from some ECU attempting to impersonate some other ECU. The purported sender $\claimed$ in that transmission corresponds to the target ECU $\target$ being impersonated. Since $\claimed$ is not transmitting at that time, the model $H_\mathrm{P}$ outputs 0, indicating that the message does not originate at $\claimed$. This contradiction reveals the presence of an attempted impersonation attack.

   \item \textbf{Detection of Compromised \acp{ecu}:} Upon detection of an intrusion and depending on the attack model, either an \ac{ecu} on the network is compromised, or an additional illegitimate module is attached to the network. If the transmission originated from a compromised \ac{ecu} $\compromised$, then the model will not predict $\claimed$ as the source of the transmission. However, there will most certainly exist an \ac{ecu} on the network for which the corresponding model will output a 1. In other words, the features of transmissions of one of the \acp{ecu} at time $u$ will closely match the previously observed features of transmissions from the corresponding \ac{ecu}.

   To detect $\compromised$, \model{} constructs the features of transmissions using power traces for all the \acp{ecu} at time $t$ except the purported source $\claimed$, that is, $\mathcal{E} \setminus \{\claimed\}$. Given the power traces of the \acp{ecu}, \model{} iterates over all the $E_k \in \mathcal{E} \setminus \{\claimed\}$ and perform model classification to determine whether transmission originated from $E_k$. The \ac{ecu} $E_k$ for which the corresponding model $H_k$ output is a $1$ is reported as the true source of the transmission $T$. This reported source \ac{ecu} is also flagged as compromised.

  \item \textbf{Detection of additional \acp{ecu}:} If the impersonating transmission originated from an additional illegitimate device that was added to the network by an attacker, then every model $H_k$ will output $0$. This indicates that none of the legitimate \acp{ecu} actually transmitted, implying the presence of an additional device on the network which sent the message.

\end{itemize}
   \subsection{\model{}-Aware Attacker}
   In this section, we argue that even if an attacker is aware of the functionality of \model{}, it is still difficult for them to mount a successful impersonation attack.
   The attacker will use their knowledge about \model{} to influence compromised \ac{ecu} power characteristics, $\bm{P}_\mathrm{C}$ to match that of the target \ac{ecu} power characteristics, $\bm{P}_\mathrm{T}$. However, the distinction between transmitting and non-transmitting states of an \ac{ecu} is most likely determined by the I/O required to transmit. And virtually all \acp{ecu} use a hardware-based \ac{can} controller to transmit, and have no physical means to transmit any other way. Thus, the attacker will be unable to do anything to cancel the inevitable power consumption profile that the \ac{can} controller exhibits when transmitting.

\section{Experimental Setup}

In this section, we give a brief overview of the setup for capturing \ac{can} transmissions and power consumption measurements from the \acp{ecu}. 
For the prototype implementation of \ac{can}, we connected four Keil MCB1700 boards to a \ac{can} bus. With each board containing a \ac{can} controller, transceiver and a receiver providing the board with the capability to send and receive \ac{can} messages with a bus speed of 125\,kbps. We supplied the boards with the same power source to ensure the minimum introduction of noise in the power consumption pattern. Using the setup, we captured a differential voltage signal from the bus and power consumption measurements from the boards using a Digitizer with a sampling rate of 10\,Msps.\par 

For the implementation of the technique in a practical setting, we deployed custom hardware in compliance with the security and warranty of the vehicle components ensuring minimal modification at the hardware level. Using the equipment, we captured \ac{can} transmissions on the bus operating at 250\,kbps, and power consumption measurements from the \acp{ecu} with a sampling rate of 10\,Msps.

\section{Results}
\label{sec:exp-evaluation}
In this section, we evaluate the proposed technique for sender authentication in a prototype setting and a sterling truck. We also show that the proposed approach applies to different bus configurations.

\subsection{Evaluation Metrics}
To evaluate the performance of \model{} for sender authentication, we report confusion matrix accuracy measures. However, different evaluation metrics examine various aspects of the experimental results. For instance, results of sender authentication for a given transmission might show varying results when evaluated on different metrics. Therefore, we assess the performance of the proposed technique for sender authentication with different bus configurations using the metrics of precision, recall, accuracy and F-measure. Precision rightly captures the false positives of the system by calculating the number of true positives (TP) over the number of true positives (TP) plus the number of false positives (FP). Whereas, recall reports the relevance of the system by calculating the number of true positives (TP) over the number of true positives (TP) plus the number of false negatives (FN). Accuracy is the ratio of correctly predicted sources for the messages to the total number of messages. And, F-measure is used to measure the similarity between the predicted state and true state by calculating the weighted average between precision and recall.

\subsection{Lab Prototype} 
With a prototype setup of the \ac{can} network, we evaluated the accuracy of \model{} for sender authentication. The transmissions in the prototype were observed from five source addresses corresponding to the five MCB1700 boards represented as five \acp{ecu}. We captured and decoded the start time and the source address of 5\,k \ac{can} transmissions observed on the bus. Using the decoded transmissions and the power consumption measurements of the \acp{ecu}, we constructed features of transmissions for the five \acp{ecu} using an estimate of the transmission window $\tau$ to be equal to 1.02\,ms. Each feature vector in the extracted features of transmissions had no of features, $M=50$, thus, generating a set of five matrices for the five \acp{ecu} of size $\numprint{25000} \times 50$. Using the mapping for the \acp{ecu} and the source addresses, we prepared the dataset for training and testing for the following combinations of \acp{ecu} and source addresses: (ECU1, SA1), (ECU2, SA2), (ECU3, SA3), (ECU4, SA4), and (ECU5, SA5) where SA1 maps to ECU1, SA2 maps to ECU, and so forth. For the mapping (ECU1, SA1), we prepared training and test set where the class of transmissions consisted of the extracted features of transmissions from \ac{ecu}1 where the source address was SA1, and the classes of non-transmission consisted of the extracted features of transmissions from \ac{ecu}1 where the source address was not SA1. Similarly, for the mapping (ECU2, SA2), the training examples with the class of transmissions comprised of features of transmissions from \ac{ecu}2 where the source address was SA2, and the examples of non-transmissions comprised of features of transmissions from \ac{ecu}2 where the source address was not SA2. Similarly, we prepared training and test sets for the rest of the source address and ECU mappings.
For model implementation, we split the training sets into balanced chunks (equal no of examples for the class of transmission and non-transmission) of training and cross-validation such that for each of the (ECU, SA) pairs, 70\% of the training examples are reserved for training and 30\% for cross-validation. Using the training and validation splits, we trained linear \acp{svm} for the five source addresses. The models finished training on the training examples with an average training time of 30 minutes. 

\subsubsection{Evaluation}
\begin{figure}[ht]
     \centering
     \includegraphics[width=0.8\linewidth]{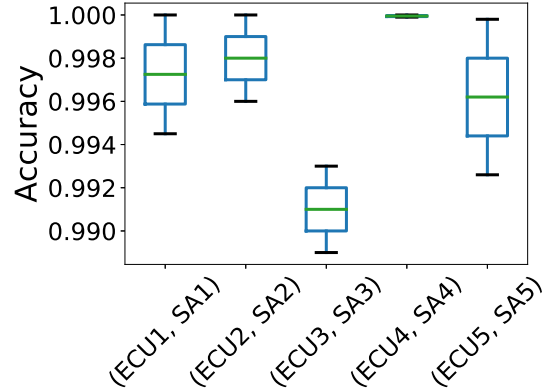}
     \caption{Models accuracies for a combination of ECUs and source addresses for the lab setup}
     \label{fig:accuracy-lab}
\end{figure}

In Figure~\ref{fig:accuracy-lab}, we report the accuracies of the five binary classifiers. The figure shows the box plots of the model accuracies obtained using bootstrapping during training and validation. Based on the figure, it is evident that the accuracy of the model for (ECU4, SA4) has relatively less variance (with less spread within and outside the box) as compared to the other models. On the other hand, the model for (ECU5, SA5) has most variance with a minimum achievable accuracy of 99.25\% and a maximum accuracy of 99.98\%. The large variance in the model accuracy is due to the presence of noise in the extracted features of transmissions. Overall, the subtle difference in the model accuracies across all the combinations shows that the technique is effective in the authentication of the sender of the message with an average accuracy of 99.87\%.  

In Figure~\ref{tab:lab}, we show the accuracy of the models to authenticate senders using the confusion matrix on the test set of 1\,k transmissions. For every transmission in the test set, we obtain a vector of the probability of transmissions from each of the models, which signifies the confidence of the models in the class of transmission. Using the probability vectors of the test set of transmissions, we obtain the confusion matrix. From the figure, we observe that the models authenticate senders of the transmissions accurately with a false positive rate of 0.0001. 

 


\begin{table*}[!ht]
\renewcommand{\arraystretch}{1.3}
\caption{Confusion matrix for sender authentication in a prototype setting}
\label{tab:lab}
\centering
\begin{tabular}{ccccccc}
\toprule
     & (ECU1, SA1) & (ECU2, SA2) & (ECU3, SA3) & (ECU4, SA4) & (ECU5, SA5)\\
\midrule
(ECU1, SA1) & \textbf{1.00} & 0.00 & 0.00 & 0.00 & 0.00\\
(ECU2, SA2) & 0.00 & \textbf{1.00} & 0.00 & 0.00 & 0.00\\
(ECU3, SA3) & 0.01 & 0.00 & \textbf{0.99} & 0.00 & 0.00\\
(ECU4, SA4) & 0.00 & 0.00 & 0.00 & \textbf{1.00} & 0.00\\
(ECU5, SA5) & 0.00 & 0.00 & 0.00 & 0.00  & \textbf{1.00}\\
\bottomrule
\end{tabular}
\end{table*}


\subsection{Real Vehicle}

We evaluated our proposed sender authentication approach in a sterling acterra truck. In the truck, we observed messages on the \ac{can} bus with three source addresses: 0, 11, and 15. Of the three source addresses, transmissions with source address 0 and 15 were observed from Engine and transmissions with source address 11 was observed from \ac{abs}. For the evaluation, we captured and decoded a total of 30\,k \ac{can} transmissions. Alongside this, we also captured the power consumption measurements of Engine and \ac{abs} from the vehicle in a stationary position. The only difference observed in the power consumption measurements of the stationary vehicle from the moving vehicle is the change in the noise floor, which the model learned to ignore over training iterations. Out of the 30\,k transmissions, 10\,k transmissions each triggered with source address 0, 11, and 15. Using the decoded transmissions and the power consumption measurements of the \acp{ecu}, we constructed features of transmissions for the three \acp{ecu} of the size $\numprint{30000} \times 50$ each using the value of $\tau$ equal to 0.65\,ms and $M=50$ principal components. Using the features, we prepared datasets for the following combinations of \acp{ecu} and source addresses: (ECM, 0), (ECM, 15), and (ABS, 11) where source address 0 and 15 are mapped to ECM (the Engine), and source address 11 is mapped to \ac{abs}. For source address 0, the class of transmissions comprised of features of transmissions from \ac{ecu} with the source address 0 and the class of non-transmission comprised of extracted features of transmissions from Engine with source addresses 11. Similarly, the dataset for source address 11 comprised of examples with the class of transmissions, which consisted of features of transmissions from \ac{abs} with source address 11 and the examples with the class of non-transmissions containing features of transmissions from \ac{abs} with source address 0 and 15. Similarly, the dataset for source address 15 is prepared.

Similar to the lab setup, we split the prepared datasets into training, cross-validation, and test set in the ratio of $6:2:2$ for the evaluation. Using the training and validation splits for the three combinations of source addresses and \acp{ecu} pairs, we implemented linear \ac{svm} using the technique described in \Section{subsec:classification}. The models finished training on the training examples with an average training time of 30 minutes.

\subsubsection{Qualitative Evaluation}

\begin{figure}[ht!]%
\centering
\subfigure[][Source Address: 0]{%
\label{subfig:sa1-tsne}%
\includegraphics[width=0.8\linewidth]{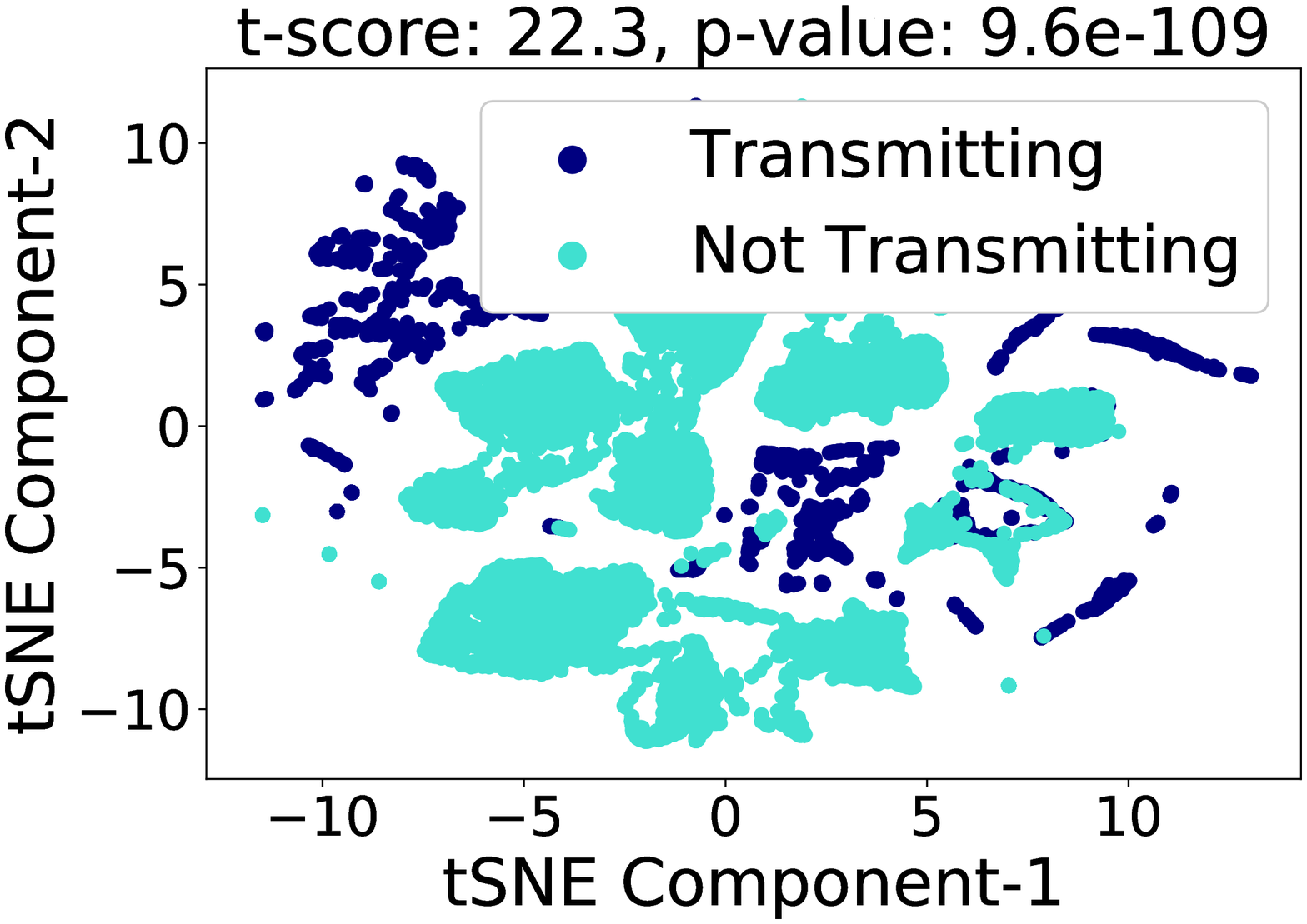}}%
\hspace{8pt}%
\subfigure[][Source Address: 15]{%
\label{subfig:sa2-tsne}%
\includegraphics[width=0.8\linewidth]{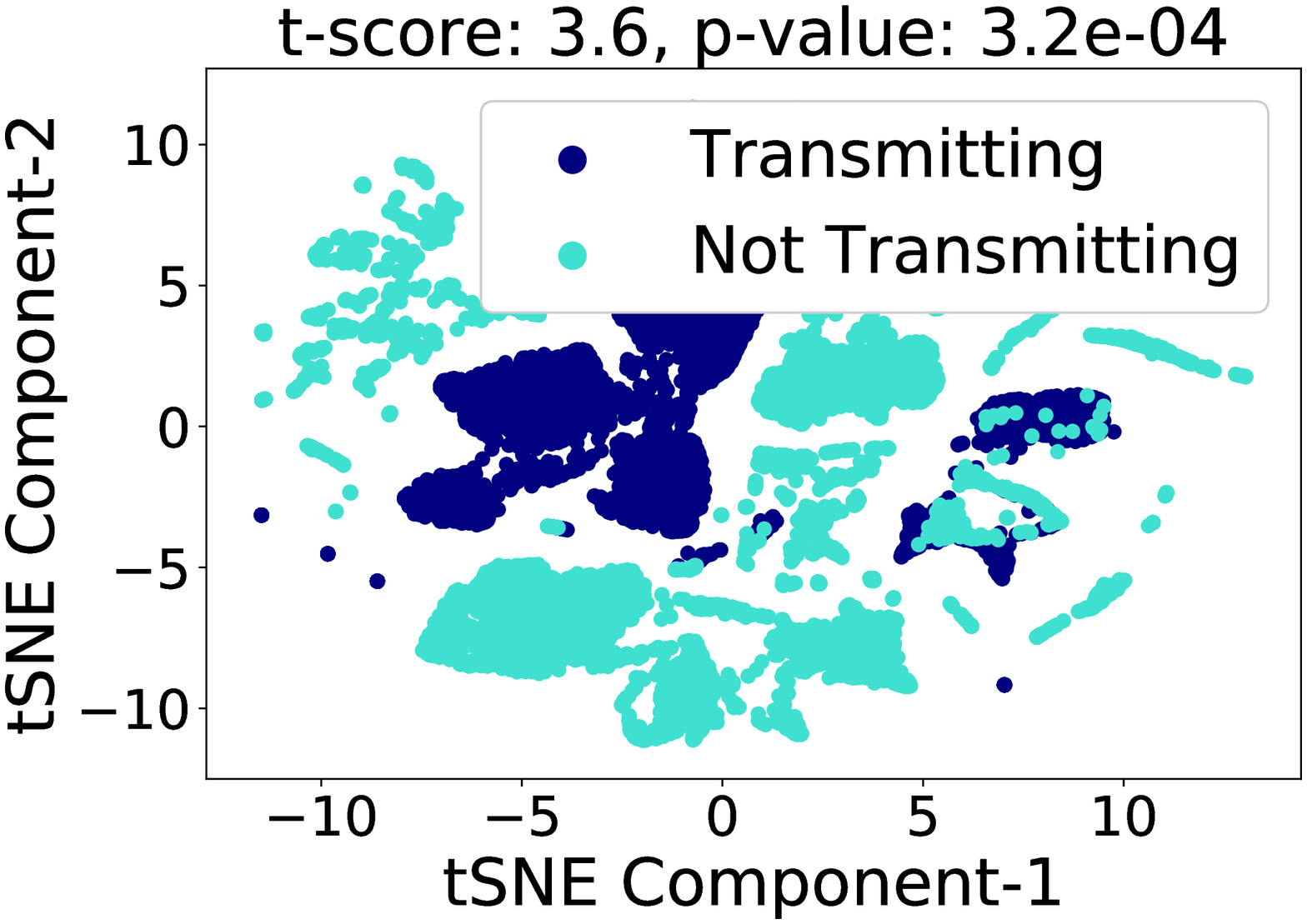}} 
\hspace{8pt}
\subfigure[][Source Address: 11]{
\label{subfig:sa3-tsne}%
\includegraphics[width=0.8\linewidth]{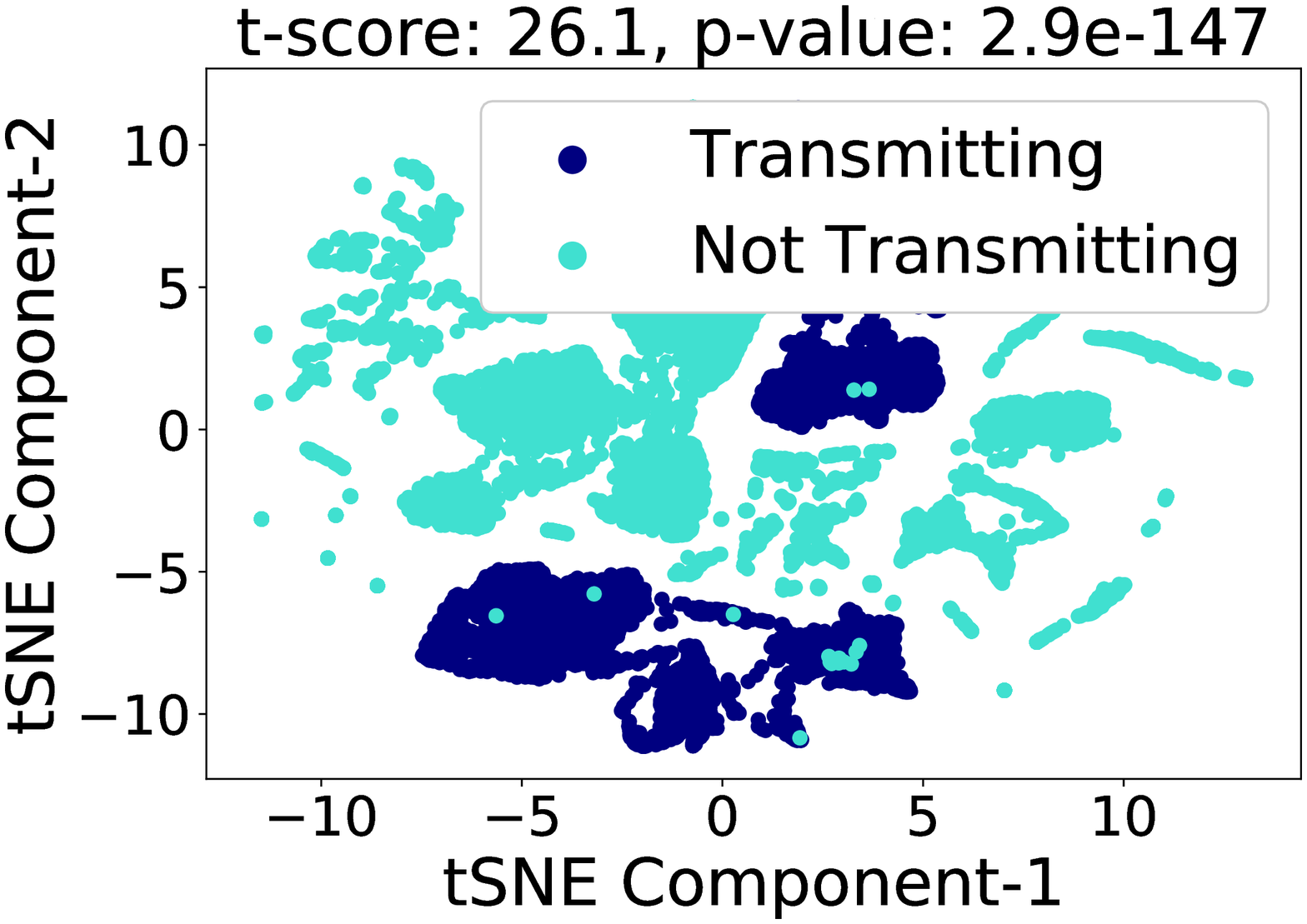}}%
\hspace{8pt}%
\caption{Scatter plots showing the tSNE components of the power traces of Engine during transmissions with source address 0, 11, and 15 in frequecy domain.}%
\label{fig:feature-tsne}%
\end{figure}

We visualized the prepared features of transmissions, and non-transmissions for the source address 0, 11, and 15. For visualization, we obtain the non-linear transformations of the features of transmissions and non-transmission for all the source addresses using tSNE. tSNE~\cite{statisticalearning} is a tool to visualize high-dimensional data by projecting to a low-dimensional non-linear subspace. \Figure{fig:feature-tsne} shows the scatter plots of the first two components of the tSNE transformations of the features for all the source addresses. The figures also show the t-scores and p-values of the features of source addresses. The t-score is a measure to tell apart the difference between the features of transmissions and the features of non-transmissions; thus, the higher the t-score value, the larger the differences. And the p-value determines the significance of t-scores. A small p-value (typically $\leqslant0.05$) provides strong evidence for the significance of the t-score. Based on the t-score and p-values of the features for the source addresses, it is evident that the characteristics of transmission are significantly different from the characteristics of non-transmissions, and hence, can be used for the identification of the state of the \acp{ecu} during a transmission.

\subsubsection{Quantitative Evaluation}

We report the accuracies of the three binary classification models using bootstrapping. Figure~\ref{fig:accuracy-lab} shows the box plots of the model accuracies. Based on the figure, it is evident that the accuracy of the model for (ECM, 0) has relatively less variance (with less spread within and outside the box) as compared to the other models. On the other hand, the model for source address 11 mapped to \ac{abs} has the most variance with a minimum achievable accuracy of 95.25\% and a maximum accuracy of 98.62\%. The large variance in the model accuracy is due to the presence of noise in the power consumption measurements of \ac{abs}, which gets propagated in the extracted features. Overall, the technique is effective in the sender authentication of \ac{can} transmissions with an average accuracy of 99.87\%.  

We evaluate the model classification accuracy on a test set of transmissions from Sterling Acterra using a confusion matrix. Similar to the lab setup, we calculate the confusion matrix using the predicted probabilities of transmissions of the models for each of the transmissions in the test set. From Table~\ref{tab:vehicle}, we observe that some transmissions with source address 0 are misclassified as 15, and vice versa resulting in a drop in the overall accuracy. This is the case because the two source address belongs to the engine \ac{ecu}, and hence, have similar power characteristics of transmissions. However, the misclassifications among source addresses, which belong to the same \ac{ecu} will still be able to identify the transmitting state of the \ac{ecu} correctly and hence, identify the sender of the message accurately. Furthermore, the response time for sender authentication in the truck is noted to be an average of 0.8\,ms, which makes the technique feasible for application in real-time settings.

\textbf{Additional Module Detection:} To test the model for the detection of an additional device (spoofed transmissions), we injected spoofed messages on the bus with the source address $0$ using a Kvaser tool~\cite{kvaser} and captured 1000 spoofed transmissions. We evaluated \model{} for sender authentication of the spoofed transmissions along with regular transmissions from the Engine and the \ac{abs} by reporting the confusion matrix in two variables: \textit{attack} and \textit{normal}. A value of one is assigned to the variable \textit{attack} if none of the source address based model's predicted probability of transmission value is greater than $\delta$, indicating that the transmission is a spoofed transmission; otherwise, a zero is assigned indicating the transmission is a regular transmission. And a value of one is assigned to \textit{normal} if atleast one of the source address based model's predicted probability of trasnmission value is greater than $\delta$ and greater than all the other model's probability value indicating the transmission is not spoofed; otherwise a zero is assigned. From the Table~\ref{tab:attack}, it is evident that all the spoofed transmissions are flagged by the technique; thus, indicating the effectiveness of the technique in sender authentication.

\begin{table}[ht]
\renewcommand{\arraystretch}{1.3}
\caption{Confusion matrix for attack detection in a Sterling Acterra Truck}
\label{tab:attack}
\centering
\begin{tabular}[t]{cccccc}
\toprule
     & \textbf{Normal} & \textbf{Attack}\\
\midrule
\textbf{Normal} & \textbf{0.99} & 0.00\\  
\textbf{Attack} & 0.00 & \textbf{1.0}\\
\bottomrule
\end{tabular}
\end{table}

\begin{figure}[ht]
     \centering
     \includegraphics[width=0.8\linewidth]{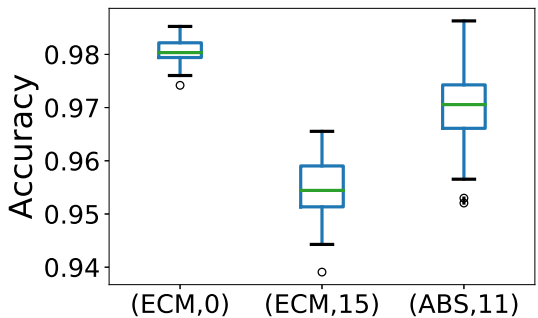}
     \caption{Accuracy of models for a combination of ECUs and source addresses}
     \label{fig:accuracy-sterling}
\end{figure}


\begin{table}[ht]
\renewcommand{\arraystretch}{1.3}
\caption{Confusion matrix for sender authentication in a Sterling Acterra Truck}
\label{tab:vehicle}
\centering
\begin{tabular}[t]{cccccc}
\toprule
     & (ECM, 0) & (ECM, 15) & (ABS, 15) \\
\midrule
(ECM, 0) & \textbf{0.99} & 0.01 & 0.00\\  
(ECM, 15) & 0.06 & \textbf{0.94} & 0.00\\
(ABS, 11) & 0.00 & 0.00 & \textbf{0.99}\\ 
\bottomrule
\end{tabular}
\end{table}

\subsection{Model Convergence}
\begin{figure}[ht!]%
\centering
\subfigure[][Source Address: 0]{%
\label{subfig:sa1}%
\includegraphics[width=0.8\linewidth]{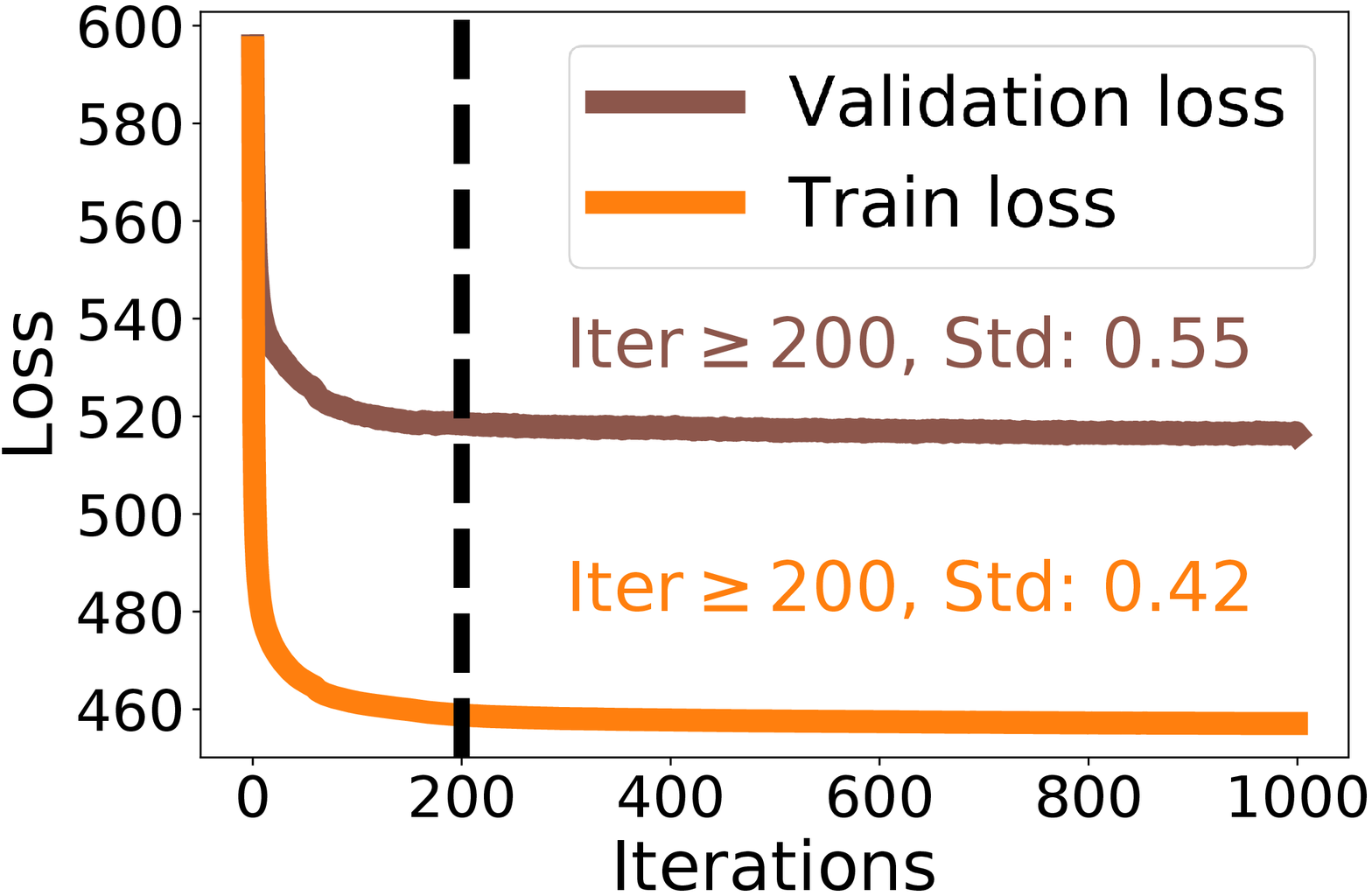}}%
\hspace{8pt}%
\subfigure[][Source Address: 15]{%
\label{subfig:comb2}%
\includegraphics[width=0.8\linewidth]{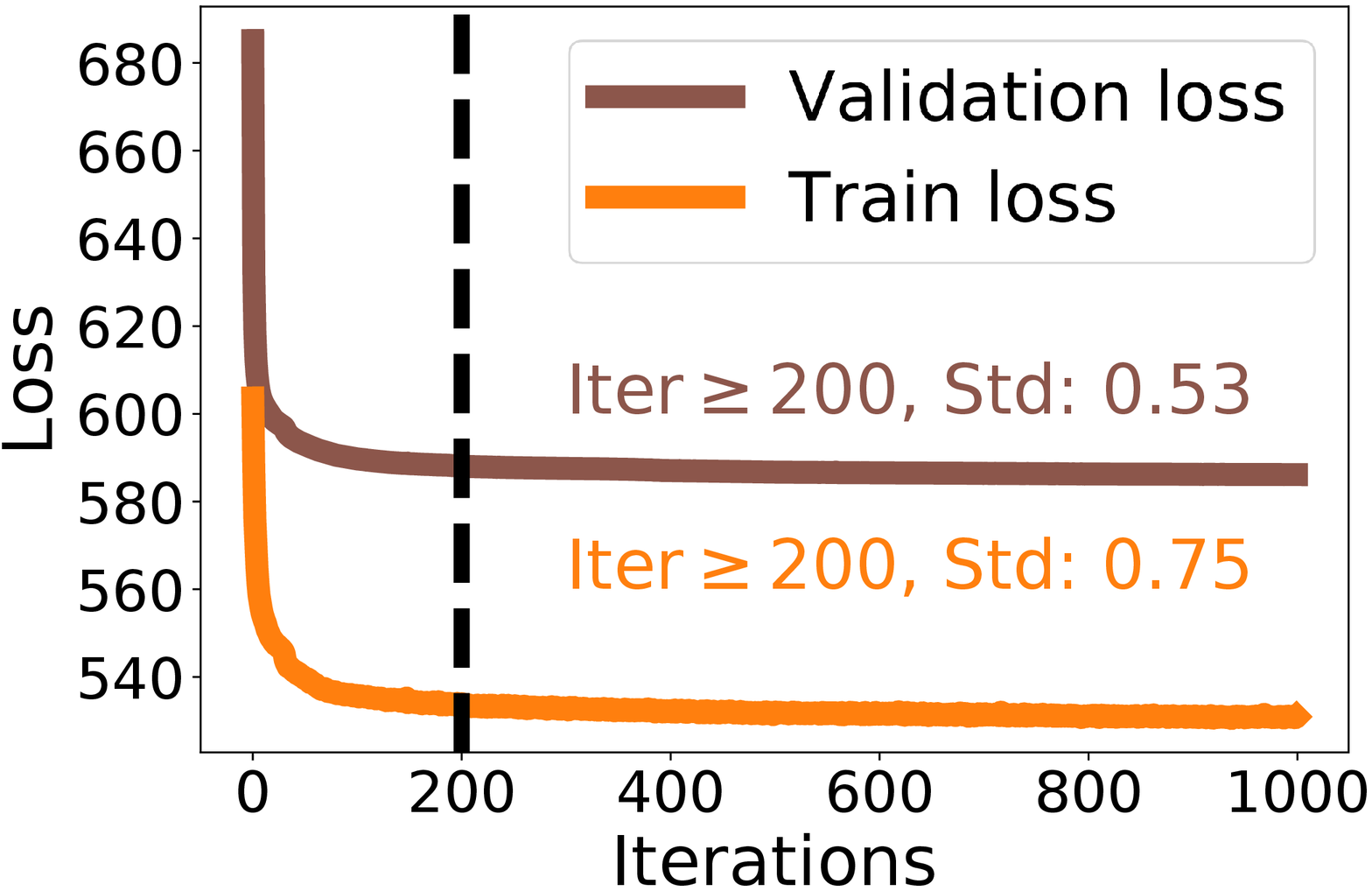}} 
\hspace{8pt}
\subfigure[][Source Address: 11]{
\label{subfig:comb3}%
\includegraphics[width=0.8\linewidth]{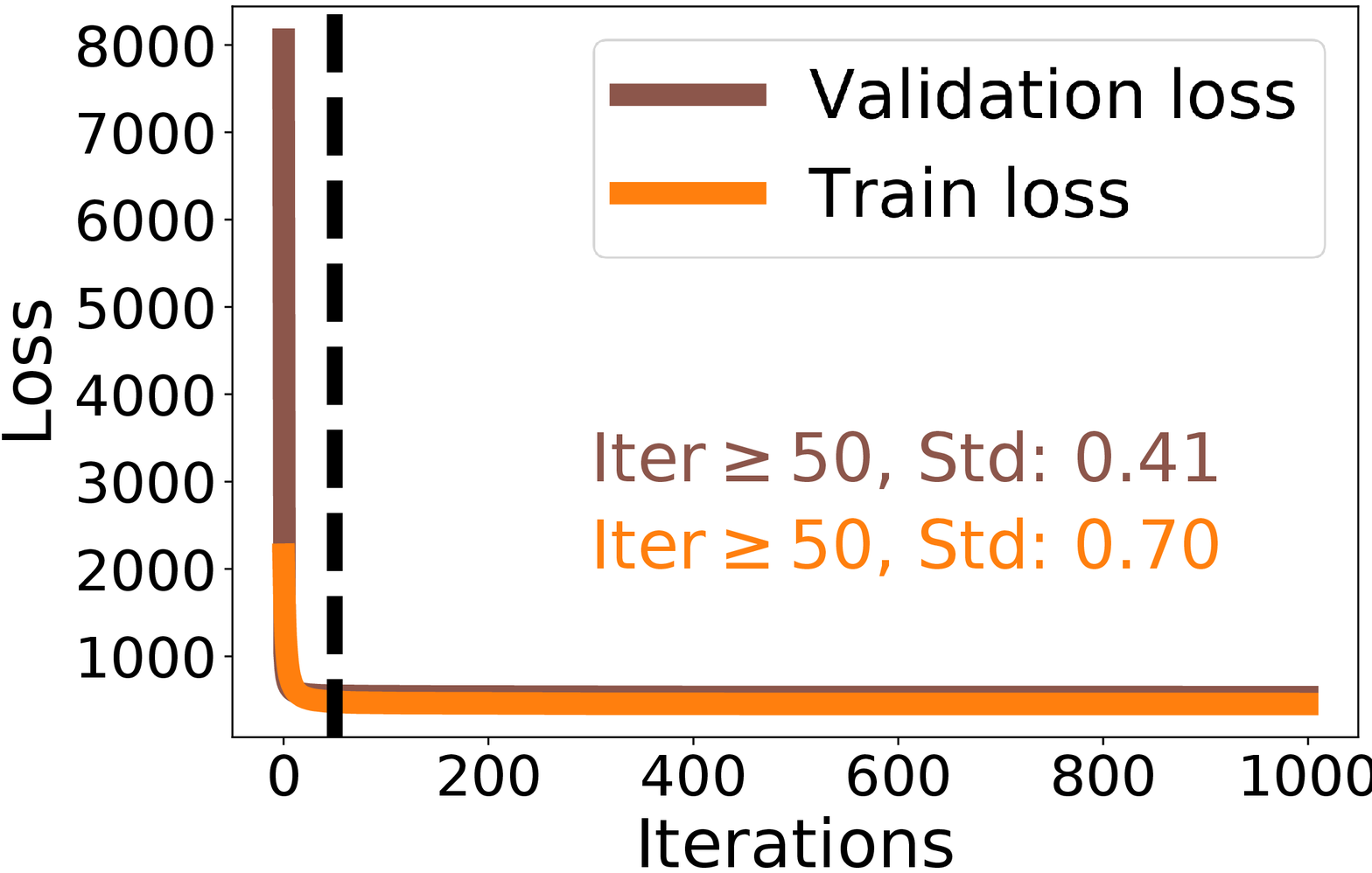}}%
\hspace{8pt}%
\caption{Plots showing the learning curves of the \ac{svm} models for sender state (transmitting/not transmitting) classification with source address 0, 11, and 15.}%
\label{fig:learning-curve}%
\end{figure}

We evaluate the convergence of the trained SVM models for sender state identification by plotting the learning curves of the models. The learning curve shows the loss (inversely proportional to the accuracy) of a model during training over many iterations. A model is said to have converged if the learning curve (including training and cross-validation curve) increases at first over the iterations and then asymptotically approaches a loss such that training any further has a negligible improvement on the performance of the model. In particular, we evaluate the learning curve of the models for the source addresses 0, 11, and 15 from real-vehicle settings. From the \Figure{fig:learning-curve}, it is evident that the model begins to converge after 200 iterations across all the source addresses. In particular, the learning curve for SA3 has a steep curve, which converges to minimum achievable loss by as early as the $50^{th}$ iteration of validation, which is shown using black dotted line after which point the standard deviation in the loss is negligible ($<1$). This is the case because, as shown in~\Figure{subfig:sa3-tsne}, the features of transmissions and non-transmissions for source address 11 form two distinct clusters in the non-linear embedding with less overlap as compared to the other two source addresses. The convergence of the model allows a model to be generalizable to unseen examples from the same distribution. Thus, the models trained with examples from sterling acterra corresponding to the transmissions captured over a timeframe of fewer than 30 minutes indicates that the technique can be applied in the truck for sender authentication of unseen transmissions without losing accuracy.

\subsection{Experimental Factors}

Power consumed by an \ac{ecu} is not immune to changes in the bus parameters and other hardware configurations. The variations introduced in the physical properties of the \ac{can} bus may manifest in the measured power consumption measurements of the \acp{ecu} resulting in power traces that deviate from the nominal power traces. The presence of variations implies that the models should be updated upon parameter changes. However, due to the complex structure of the modern automobile system, it is not feasible to update models upon every hardware or firmware update. Therefore, we study the impact of the variations in parameters such as bus speed, message format and embedded programs on the power consumption measurements of the \acp{ecu}. 

\par
The factors and the corresponding levels against which we evaluate the approach are as follows. Bus speed: 125\,kbps, 250\,kbps, and 500\,kbps, Message format: standard (11-bit) and extended (29-bit), embedded programs: uniform and heterogeneous. The selected levels of the bus configuration are found in the majority of the modern automobile with 125\,kbps used for low-speed \ac{can} communications and 500\,kbps for high-speed communications \cite{CAN}. The two levels of the message format are the only possible format of \ac{can} frames in a \ac{can} protocol. At the uniform level of the source code, we executed the same program before and after all the transmissions in the source \ac{ecu}. Whereas, at the heterogeneous level, we executed a randomly picked source code from amongst a suite of mibench source code \cite{mibench} before and after the transmissions in the source \ac{ecu}. 

\begin{figure*}[ht]%
\centering
\includegraphics[width=0.9\linewidth]{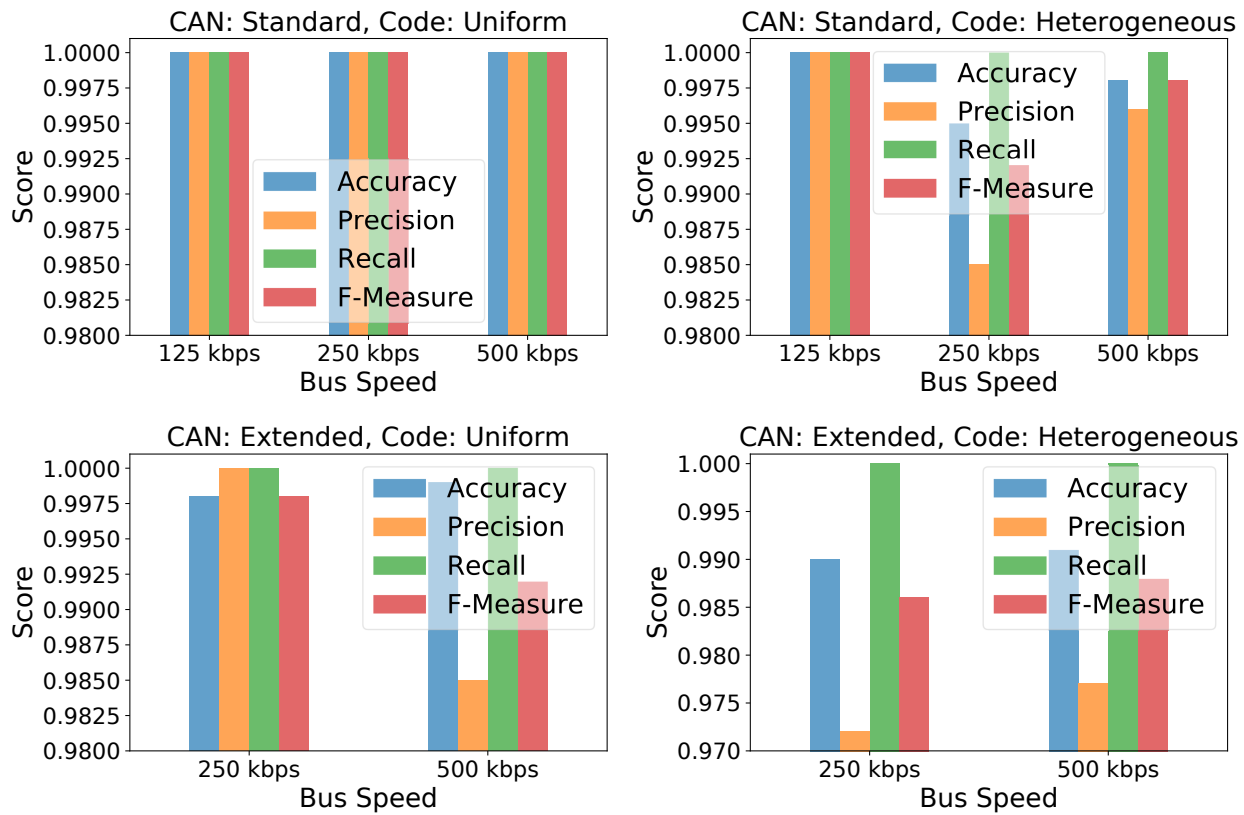}%
\caption{Figure shows the impact of variation in the bus speed, \ac{can} frame format, and source code on the accuracy of \model{}. For each combination of the experimental factors, the subplots show the metrics of accuracy, precision, recall, and F-measure.}%
\label{fig:experimental-factors}%
\end{figure*}

We identified the critical interaction between factors and the corresponding levels using fractional factorial design analysis \cite{montgomery}. Based on the analysis of variance experiment, four of the 25 interactions are reported as most significant. Using the significant interactions in the lab setup, we conducted the experiments in the lab setup. For every experiment, we captured and decoded 5\,k transmissions. For the decoded transmissions, we constructed features of transmissions using the proposed approach~\ref{subsec:features-transmissions} for all the source addresses mapped to the \acp{ecu} across all the experiments. We trained, cross-validated and evaluated the \acp{svm} for each of the experiments using the train, cross-validation and test set of power traces. \Figure{fig:experimental-factors} shows the results of classification. Results show that the proposed technique is more accurate with simple bus configuration (125\,kbps bus speed, Standard Format, Uniform Source Code) than with advanced configurations (500\,kbps bus speed, Extended Format, Heterogeneous Source Code). 
However, the subtle difference between the accuracy of the two extreme network configurations shows that the impact of the variations of the factors has a negligible effect on the state classification accuracy of the \ac{ecu}. And hence, the proposed approach can be applied to practical settings without worrying about the impact of small and frequent changes in the bus properties.


\section{Discussion}
\label{sec:discussion}
We use non-clonable characteristics of the transmission to develop a sender authentication technique. 
An important and unique advantage of our proposed technique is that \model{} relies on \emph{truly non-clonable} characteristics of the transmitter. Previous works based on physical characteristics rely on attributes of the transmitted signal, which in principle can be imitated. In our case, we rely on physical attributes \emph{of the transmitter}. This means that under the assumption that \model{}'s implementation is secure and tamper-proof, then even with unrestricted physical access to the \ac{can} bus, an attacker is still physically unable to defeat \model{}.

\subsection{Applications}
The applicability of \model{} is not limited to \ac{can} protocol but to many other communication protocols with similar properties, notably, lack of and difficulty to incorporate sender authentication mechanisms. In particular, \model{} can also be used with a more refined version of \ac{can} protocols such as FlexRay, TTCAN, CAN-FD, CANopen, and SafetyBUS. Moreover, the fact that these protocols are next-generation protocols implies that \model{} applies to future generation automobile systems as well. 

\subsection{Future Work}
Avenues for future work on \model{} include improvement of the technique in its current form as well as extending the idea for a more flexible and generalizable sender authentication mechanism.
In the current implementation of \model{}, we use the power consumption measurements from the \acp{ecu} for sender authentication. An interesting extension to \model{} would be to use the power consumption of the entire system instead to isolate the transmitting \ac{ecu}. The approach helps reduce the number of hardware interfaces to monitor power consumption measurements from \acp{ecu} as well as the computational complexity. Furthermore, as \model{} relies on the verdict of the model to report the true state of any \ac{ecu}, it is essential to determine what aspects of the input power traces triggers the model to conclude. An interesting future work as an enhancement of the proposed technique is to unravel the root cause of the model output.

\section{Conclusion}
\label{sec:conclusion}

 In this paper, we proposed a novel technique for sender authentication using power consumption measurements of the \acp{ecu}. The capability of the power trace based method to accurately determine the transmitting state of the \acp{ecu} result in accurate sender identification. We showed that the approach can be used to detect the presence of compromised and additional devices on the network. Evaluation of the approach against a lab setup and a practical setting showed that the technique is highly effective with a false positive rate of 0.002\%. We also showed that the approach applies to different network settings without compromising on the accuracy.
\bibliographystyle{IEEEtran}
\bibliography{emsoft2020}
\end{document}